\documentclass[aps,twocolumn,noshowpacs,superscriptaddress,floatfix]{revtex4}

\usepackage[utf8]{inputenc}
\usepackage{braket}
\usepackage{amsmath,amssymb}
\usepackage[justification=justified]{caption}
\usepackage{qcircuit}
\usepackage{xcolor}
\usepackage{verbatim}
\usepackage[normalem]{ulem}
\usepackage[rightcaption]{sidecap}

\usepackage{graphicx}
\usepackage{subfig}

\usepackage[section]{algorithm} 
\usepackage{amsthm,algorithmic,epsfig,tikz}

\newcommand{\R}{\mathbb{R}}

\def\bra#1{\mathinner{\langle{#1}|}}
\def\ket#1{\mathinner{|{#1}\rangle}}

\renewcommand{\part}[2]{\frac{\partial #1}{\partial #2}}

\newcommand{\al}[1]{\begin{align} #1\end{align}}

\usepackage{float}
\usepackage{multirow}

\begin{document}

\title{Low depth amplitude estimation on a trapped ion quantum computer}

\author{Tudor Giurgica-Tiron}
\affiliation{Goldman, Sachs \& Co.}
\affiliation{Stanford University, Palo Alto, CA.}

\author{Sonika Johri}
\affiliation{IonQ Inc, 4505 Campus Dr, College Park, MD 20740}

\author{Iordanis Kerenidis}
\affiliation{QC Ware, Palo Alto, USA and Paris, France}
\affiliation{IRIF, CNRS - University of Paris, France}

\author{Jason Nguyen}
\affiliation{IonQ Inc, 4505 Campus Dr, College Park, MD 20740}

\author{Neal Pisenti}
\affiliation{IonQ Inc, 4505 Campus Dr, College Park, MD 20740}

\author{Anupam Prakash}
\affiliation{QC Ware, Palo Alto, USA and Paris, France}

\author{Ksenia Sosnova}
\affiliation{IonQ Inc, 4505 Campus Dr, College Park, MD 20740}

\author{Ken Wright}
\affiliation{IonQ Inc, 4505 Campus Dr, College Park, MD 20740}

\author{William Zeng} 
\affiliation{Goldman, Sachs \& Co.}

\date{\today}

\begin{abstract} 

Amplitude estimation is a fundamental quantum algorithmic primitive that 
enables quantum computers to achieve quadratic speedups for a large class of statistical estimation problems, including Monte Carlo methods. The main drawback from the perspective of near term hardware implementations 
is that the amplitude estimation algorithm requires very deep quantum circuits. Recent works have succeeded in somewhat reducing the necessary resources for such algorithms, by trading off some of the speedup for lower depth circuits, but high quality qubits are still needed for demonstrating such algorithms. 

Here, we report the results of an experimental demonstration of amplitude estimation on a state-of-the-art trapped ion quantum computer. The amplitude estimation algorithms were used to estimate the inner product of randomly chosen four-dimensional unit vectors, and were based on the maximum likelihood estimation (MLE) and the Chinese remainder theorem (CRT) techniques. Significant improvements in accuracy were observed for the MLE based approach when deeper quantum circuits were taken into account, including circuits with more than ninety two-qubit gates and depth sixty, achieving a mean additive estimation error on the order of $10^{-2}$. The CRT based approach was found to provide accurate estimates for many of the data points but was less robust against noise on average. Last, we analyze two more amplitude estimation algorithms that take into account the specifics of the hardware noise to further improve the results.

\end{abstract} 

\maketitle

\section{Introduction}
Amplitude estimation \cite{BHMT02} is a fundamental quantum algorithmic primitive that 
enables quantum computers to achieve quadratic speedups for a large class of statistical estimation problems. 
Amplitude estimation is of particular interest for quantum algorithms in computational finance 
as it underlies the quantum speedup for Monte Carlo methods \cite{M15}. 

In a typical setting for quantum Monte Carlo methods, the quantum algorithm has access to an evaluation oracle 
for a function $f: X \to \R$ and the goal is to estimate the mean value of $f$ to additive error $\epsilon$. 
Montanaro \cite{M15} gave an algorithm to estimate the mean for arbitrary distributions on $X$ in the setting where 
the variance of $f$ is upper bounded by $\sigma$. It requires  $O(\sigma/\epsilon)$ calls to the oracle for 
$f$, achieving a quadratic speedup over classical algorithms that require $O(\sigma^{2}/\epsilon^{2})$ calls. 
Quantum Monte Carlo methods have also been developed for settings where the 
distribution on $X$ is uniform \cite{A99,G98,B11} and for settings with known bounds on the mean \cite{L18,H18}. 

The main drawback of quantum Monte Carlo methods from the perspective of near term hardware implementations 
is that the amplitude estimation algorithm requires the evaluation oracle for $f$ to be invoked 
sequentially $O(\sigma/\epsilon)$ times. The depth of the quantum circuit for Monte Carlo is therefore 
$O(\sigma D(f)/\epsilon )$ where $D(f)$ is the depth of the evaluation oracle. Reducing the resource 
requirements for quantum Monte Carlo methods to bring them closer to implementation on noisy 
intermediate scale quantum (NISQ) architectures has been the focus of a number of recent works. 

The first results on reducing resource requirements for amplitude estimation were given by Suzuki \emph{et al.} 
\cite{S20} who used maximum likelihood estimation with exponential schedules to eliminate the quantum Fourier transform (QFT) step in the amplitude estimation algorithm. Aaronson and Rall \cite{AR20} gave a QFT free algorithm with rigorous guarantees and further work by Grinko \emph{et al.} \cite{grinko2019iterative} on iterative AE reduced the constant overheads for the QFT free amplitude estimation. 
Eliminating the QFT does not lower significantly the depth of the AE algorithm as the QFT circuit has depth $O(\log \log(1/\epsilon))$, but it is an important step towards making amplitude estimation nearer term. The QFT free algorithms do not need to apply controlled versions of the oracle. 

A different line of research, aimed at reducing the depth for AE algorithms while still retaining a partial speedup, was initiated in \cite{GKLPZ20} where AE algorithms with depth $O(1/\epsilon^{1-\beta})$ and $O(1/\epsilon^{1+\beta})$ queries were proposed and analyzed. These algorithms interpolate between classical sampling and the standard AE algorithms and are based on maximum likelihood estimation with power law schedules and boosting AE estimates using the Chinese remainder theorem. Similar tradeoffs for the special case of approximate counting were obtained by Burchard \cite{B19} with supporting results from the lower bound literature indicating the optimality of the tradeoff \cite{J17, Z99}. Tanaka \emph{et al.} \cite{TS20} studied amplitude estimation in the presence of depolarizing noise and gave evidence that super linear power law schedules can be robust against noise. More recent work has proposed a Fourier series decomposition to reduce the amplitude estimation depth \cite{herbert2021quantum} and a variational approach to near term term amplitude estimation \cite{plekhanov2021variational}.

Applications of quantum Monte Carlo methods to problems in computational finance has been the focus of several recent works. These works explore the application of quantum Monte Carlo to specific financial problems like pricing simple options and credit risk calculations 
\cite{R18, egger2019credit, W19, miyamoto2021bermudan, S19}. Some of these works have an experimental component, for example  \cite{S19, W19} report proof-of-concept results of running their models on three and four qubit quantum devices. Experimental results are limited both by the number of qubits available and the circuit depth of the amplitude estimation procedure, so a quantum advantage over classical Monte Carlo methods has not been established so far due to hardware limitations. 
The costs for implementing the amplitude estimation oracle for these applications are significant. Therefore, some works have focused on resource estimation \cite{S19, GS} 
while others have tried to reduce the cost of implementing the oracles for financial applications \cite{miyamoto2019reduction, ramos2019quantum,vazquez2020efficient,kaneko2020quantum}. 

Experiments on near term amplitude estimation using the iterative AE \cite{grinko2019iterative} and MLE based approaches \cite{S20} for two and three qubits were recently carried out on the IBM quantum computers \cite{rao2020quantum}. Amplitude estimation was used for Monte Carlo estimation of the integral $\int^{\pi/2}_{0}\sin^{2} (x) dx$. The main evaluation oracle circuit consisted of three qubits and two two-qubit gates, the maximum depth that was demonstrated was two, and the hardware noise led to significant loss of performance for the AE algorithm compared to simulator runs. The additive error for estimating the integral was also fairly large compared to the magnitude of the integral, thus attesting to the limitations of the hardware. 

Noise aware approaches to amplitude estimation that consider a Gaussian noise model instead of the depolarizing channel and adapt the AE algorithm to the noise have been proposed very recently \cite{herbert2021noise}. Experimental results on the Honeywell quantum hardware show that incorporating information about the noise can improve performance of the AE algorithms. 


In this paper, we report the results of experiments demonstrating the low depth AE algorithms from \cite{GKLPZ20} on IonQ's state-of-the-art trapped ion quantum computer.
We compare the AE performance against sampling directly from the evaluation oracle. A successful demonstration shows that AE has less estimation error for the same number of samples. On a noiseless quantum computer the advantage from the AE approach would be quadratic in the depth of the quantum algorithm run. As current hardware has noise, we expect to see the AE advantage increase to some maximum depth after which noise from longer circuits overwhelms the quantum scaling advantage. This demonstration on IonQ hardware (Fig. \ref{fig:MLE}) is the main result of this work.

Note that this is a proof-of-concept demonstration of the scaling advantage of AE. This is because the classical sampling comparison we use still uses the evaluation oracle on the QPU. To show practical quantum advantage, the quantum AE approach would need to best classical sampling with classical processors that today have a much higher clock speed and negligible error rates compared to QPUs. Significant improvements in QPU performance will be needed before this demonstration of practical advantage.

Here, amplitude estimation is used to provide more accurate estimates of the inner product between randomly chosen four-dimensional vectors. The circuits for inner product estimation have been developed using quantum data loaders, which are logarithmic depth quantum circuits for creating vector states. The data loaders were introduced in \cite{johri2021nearest}, where they were used for state preparation for quantum nearest centroid classification of the MNIST dataset, and have also being used for loading data in quantum neural networks \cite{qnn2021, qorthonn2021}. 
The inner product estimation circuits, used as the evaluation oracle for amplitude estimation, when compiled down to two-qubit gates that can be directly applied on the hardware, are circuits on four qubits that use eight two-qubit gates and have depth six. 

The amplitude estimation algorithm executes these oracles in sequence a number of times in order to get more accurate estimates of the inner product. The high gate fidelity for the trapped ion quantum computer allow us to report experimental results for considerably deep amplitude estimation circuits, where the oracle circuit is repeated sequentially up to fifteen times for a total depth of more than sixty and a total number of two-qubit gates of more than ninety.  

We use two different low-depth amplitude estimation algorithms, one based on a maximum likelihood estimation (MLE) applied on the results of different depth circuits, and the second is an adaptation of the QoPrime algorithm from \cite{GKLPZ20}, where quantum circuits for two co-prime depths are run. 

The main figure of merit of whether the amplitude estimation algorithms have been demonstrated in practice is whether the accuracy of the estimation when using samples from higher depth quantum circuits (i.e. applying the evaluation oracle sequentially many times) is better than the accuracy achieved when one only uses samples from the evaluation oracle directly. Note that there is a level of noise in the circuits which has some interesting consequences: first, samples beyond a certain depth are going to be ineffective at improving further the accuracy, since they become too noisy; and second, even for sampling from the evaluation oracle directly, a certain approximation error will remain no matter if we increase the number of samples. We explain these effects in more detail in the following sections.   

Let us provide some high level description of the results of the demonstration. The results of the MLE based AE with higher depths demonstrate significant improvements in the accuracy compared to using a large number of samples from the evaluation oracle. Here we use a simple linear schedule, thus considering circuits of $2t+1$ sequential oracle calls for $t$ all possible integers up to 7. These results constitute a proof-of concept demonstration of MLE based amplitude estimation on real quantum hardware. We note that for such small depths it does not make much sense to consider the more involved power law schedules based AE as described in \cite{GKLPZ20}, which will start providing better tradeoffs between the total number of oracle calls and the circuit depth as the hardware continues to evolve. 

Compared to the MLE based approach, the QoPrime AE algorithm, based on the Chinese remainder theorem, is less robust to noise, though it has a rigorous proof of correctness. We report experimental results for CRT based algorithms where the moduli are chosen to be consecutive odd numbers. The results on CRT based AE show convergence for the pair of moduli (3,5), for other pairs of moduli the algorithm converges on a large fraction the data points but not on all. The reason behind this performance is that the theoretical analysis of the CRT algorithm assumes that the sampling probabilities behave like independent coin flips and applies Chernoff bounds to bound the resulting errors. However, in practice, the errors on consecutive shots may be correlated. This results in some of the data points having inaccurate estimates, leading to higher average errors for the CRT based algorithm. Nevertheless, the CRT based algorithm in the noiseless case is more accurate than the MLE algorithm, and, therefore, it has the potential to be competitive with the MLE based approach with further improvements in hardware fidelity and calibration. 

Last, we analyze two more algorithms that take advantage of knowledge of the noise present in the hardware. This information may come from the calibration of the hardware directly or better by running the algorithms on some training data beforehand, calculating the appropriate hyper parameters for the specific hardware, and then running the amplitude estimation algorithms on the desired inputs with those hyper parameters.
The first algorithm is a hybrid algorithm that chooses between an MLE based estimate for very small depths and a CRT based estimate for two large moduli, and whose test condition is tuned experimentally.
The second such algorithm is an MLE based amplitude estimation algorithm with a power law schedule, as defined in \cite{GKLPZ20}, where the exponent is tuned experimentally.

In the following sections, we describe in detail the amplitude estimation algorithms and circuits and provide an analysis of the results of the experimental demonstration of the different AE algorithms.

\section{The amplitude estimation algorithm and circuits}

We consider an amplitude estimation setting where the algorithm is given access to a quantum circuit $\mathcal{A}$ such that 
$\mathcal{A}\ket{0^k} =  \sin(\theta) \ket{1, x} + \cos (\theta) \ket{0, x'} $ where $\ket{x}, \ket{x'}$ are arbitrary states on $(k-1)$ qubits. The algorithm's goal
is to estimate the angle $\theta$ within an additive factor $\epsilon$. We call $\mathcal{A}$ the evaluation oracle. 

The iteration circuit consists of the following four unitaries $ \mathcal{A} S_0 \mathcal{A^\dagger} S_\chi$, where the reflection $S_\chi$ puts a minus phase in front of the states $\ket{1}\ket{x}$ and $S_0= 2\ket{0^n}\bra{0^n}-I$ is a reflection around the all-zero state. 

The low-depth quantum amplitude estimation algorithm executes a number of different circuits $U^t$, for different depths $t$, where we define these circuits as
\begin{equation}
\label{Ut}
U^t = (\mathcal{A} S_0 \mathcal{A^\dagger} S_\chi)^t \mathcal{A}
\end{equation}

In other words, the depth 0 circuit is a simple application of the evaluation oracle $\mathcal{A}$, while for a depth $t$ we add $t$ applications of the iteration circuit to an initial application of the oracle. Thus, for depth $t$ the number of oracle calls is $(2t+1)$ and the state at the end of the circuit $U^t$ is

\begin{equation*}
U^t \ket{0} =  \sin((2t+1)\theta) \ket{1, x} + \cos ((2t+1)\theta) \ket{0, x'}
\end{equation*}

We now describe the evaluation oracle that will be used as our amplitude estimation benchmark. We use a simple circuit for estimating the square inner product between two random unit vectors that are encoded with a unary amplitude encoding method. Such circuits have been used before for estimating the Euclidean distance between data points in \cite{johri2021nearest}.
In Figure \ref{fig:oracle} we see the inner product estimation circuit for 4-dimensional vectors that uses four qubits and four $RBS$ gates (denoted as B-S in the circuit). 

\begin{figure}[h]
    \centering
    \includegraphics[width=70px]{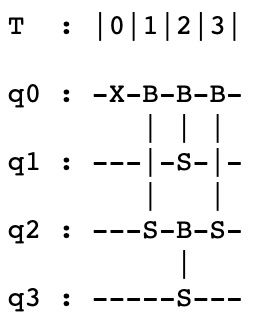}
    \caption{The inner product estimation oracle for 4-dimensional inputs. It consists of an initial $X$ gate and four $RBS$ gates whose parameters depend on the input vectors.}
    \label{fig:oracle}
\end{figure}

An $RBS(\phi)$ gate is a two qubit gate defined as
\begin{equation} \label{RBS}
RBS(\phi) = \left( \begin{array}{cccc}
1 & 0 & 0 & 0 \\
0 & \cos \phi & \sin \phi & 0 \\
0 & -\sin\phi & \cos\phi & 0 \\
0 & 0 & 0 & 1  \end{array} \right)
\end{equation} 
Note also that $RBS^\dagger(\theta) = RBS(-\theta)$.

Given two unit vectors $x$ and $y$, one can find angles $\phi$ for the four $RBS$ gates in the circuit so that the final state at the end of the quantum circuit is of the form 
\begin{equation}
x\cdot y \ket{e_1} + \ket{G}
\end{equation}
where $\ket{G}$ is an unnormalised unary state orthogonal to $e_1$, in other words whose first qubit is in state $\ket{0}$. This implies that when we measure the first qubit of this quantum state, the probability we get outcome $1$ is exactly $(x\cdot y)^2$, while the outcome is 0 with the remaining probability $1-(x\cdot y)^2$. In order to match the notation for the amplitude estimation algorithm, one can rewrite the above state as
\begin{equation}
\sin \theta \ket{1} \ket{000} + \cos \theta \ket{0}\ket{000^\bot}
\end{equation}
for $\theta$ such that $\sin \theta = x\cdot y$.
Hence, we can use this inner product estimation circuits as the evaluation oracle for an amplitude estimation implementation.
We see that in total the evaluation oracle consists of four $RBS$ gates with a circuit depth of three. Here we focus on the number and depth of the two-qubit gates, since single-qubit gates are much faster and easier to implement.   

\begin{figure}[h]
    \centering
    \includegraphics[width=250px]{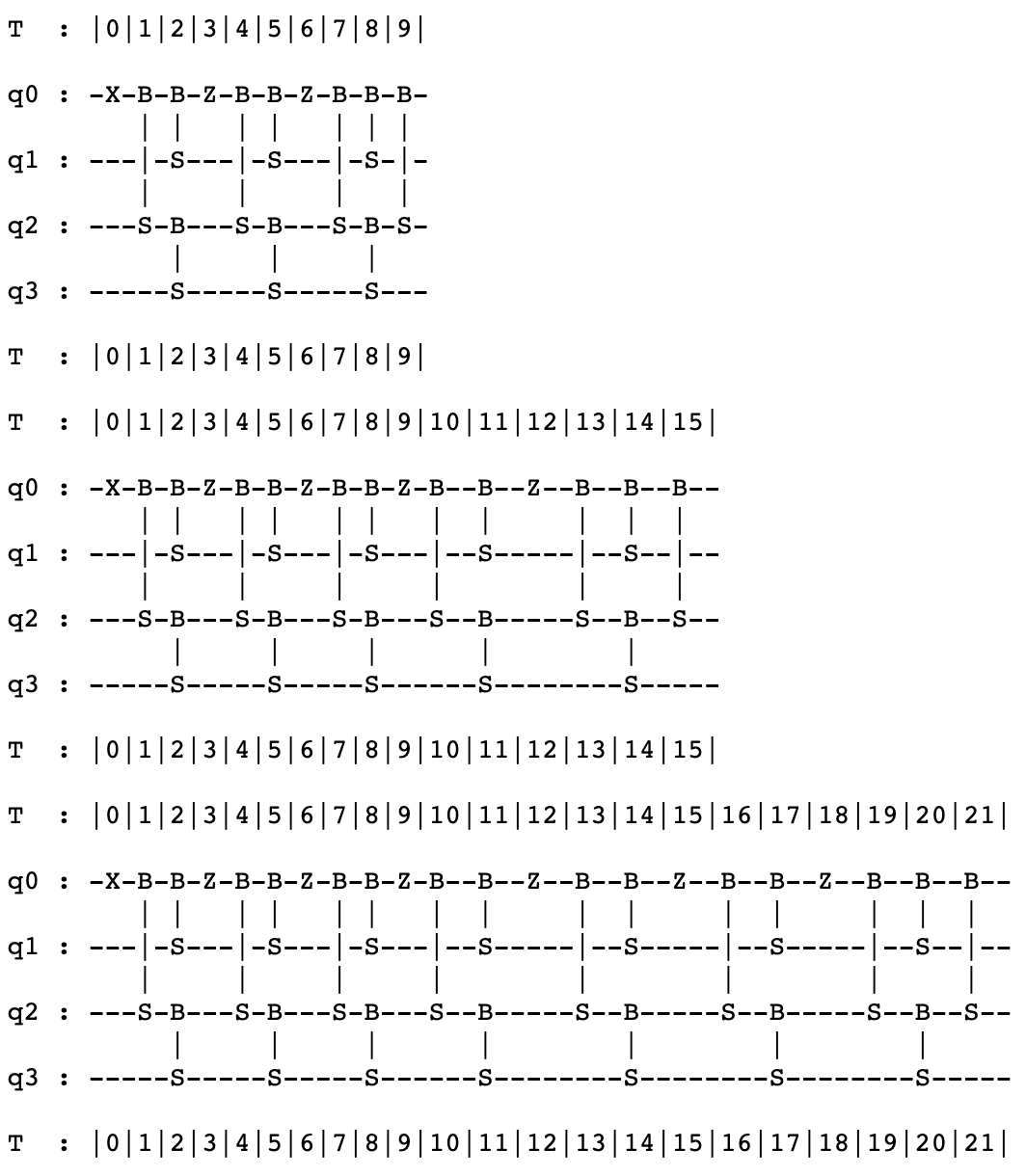}
    \caption{The circuits $U^t$, for depth $t=1,2,3$. The hardware implementation consisted of circuits up to $t=7$, with a total of ninety two two-qubit gates and a circuit depth of sixty two.}
    \label{fig:final}
\end{figure}

We remark here that one could use many different evaluation oracle circuits in order to test the amplitude estimation algorithms. The inner product estimation circuits used in the experiments as the evaluation oracle are known to be optimal both with respect to the number of gates and circuit depth and they have a number of different applications in estimating the Euclidean distance between data points in similarity-based classification algorithms \cite{johri2021nearest} or for assisting in the training of neural networks \cite{qnn2021, fnn2020, cnn2019}. They also provide a great benchmark for the AE algorithms, since one can easily increase the complexity of the oracle by increasing the dimension of the unit vectors. For example, using 8-dimensional vectors increases the complexity of the evaluation oracle to a circuit with eight qubits, ten $RBS$ gates and depth five.  

We will now see how to optimize the circuits of the amplitude estimation before moving to the real hardware implementation.
First, we define the operators in the iteration circuit. For implementing $\mathcal{A}^\dagger$, one only needs to invert the circuit and negate the angles of the $RBS$ gates.
The fact of using a unary encoding also greatly simplifies the implementation of the two reflections, in fact it is easy to see that the reflections both simplify to adding a single $Z$ gate on the first qubit between applications of the oracle and its adjoint.

The last optimization of the circuit is to notice that two $RBS$ gates with a $Z$ gate between them can be simplified to only one $RBS$ gate and one $Z$ gate, by changing the angle parameter of the $RBS$ gate. This simplification happens in all applications of the oracle apart from the last one. Thus, the final circuits we implement appear in Figure \ref{fig:final}, where we show the circuits for depth 1,2, and 3. Note that we implemented circuits up to depth 8.

\begin{figure}[h]
    \centering
    \includegraphics[width=100px]{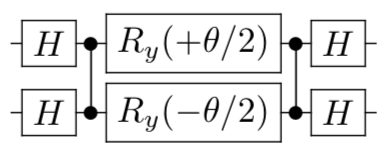}
    \caption{A decomposition of the $RBS(\theta)$ using Hadadard gates $H$, two single qubit rotations $R_y$ and two two-qubit $CZ gates$.}
    \label{fig:RBS_implementation}
\end{figure}

Last, we use the decomposition of the $RBS$ gate into single-qubit gates and two $CZ$ gates as in Figure \ref{fig:RBS_implementation} \cite{qnn2021} in order to implement it efficiently with the two-qubit gates available on the IonQ platform.  

Hence, for the circuit $U^t$, the number of two-qubit gates implemented on the hardware are $(12t+8)$ for a circuit depth of $(8t+6)$.




\section{Experimental Apparatus}
The experimental demonstration was performed on the newest generation IonQ quantum processing unit (QPU). This system, as in previous IonQ QPUs \cite{benchmarking2019}, utilizes trapped Ytterbium ions where two states in the ground hyperfine manifold are used as qubit states. These states are manipulated by illuminating individual ions with pulses of 355 nm light that drive Raman transitions between the ground states defining the qubit. By configuring these pulses, arbitrary single qubit gates and Molmer-Sorenson type two-qubit gates can both be realized. This QPU features not only an order of magnitude better performance in terms of fidelity but also is considerably more robust compared to its current QPU on the cloud. This allows for deep circuits with many shots to be run over a very reasonable period of time. This increased data collection rate has made it possible to run experiments of this nature. 

\begin{figure}[h]
    \centering
    \includegraphics[width=0.4\textwidth]{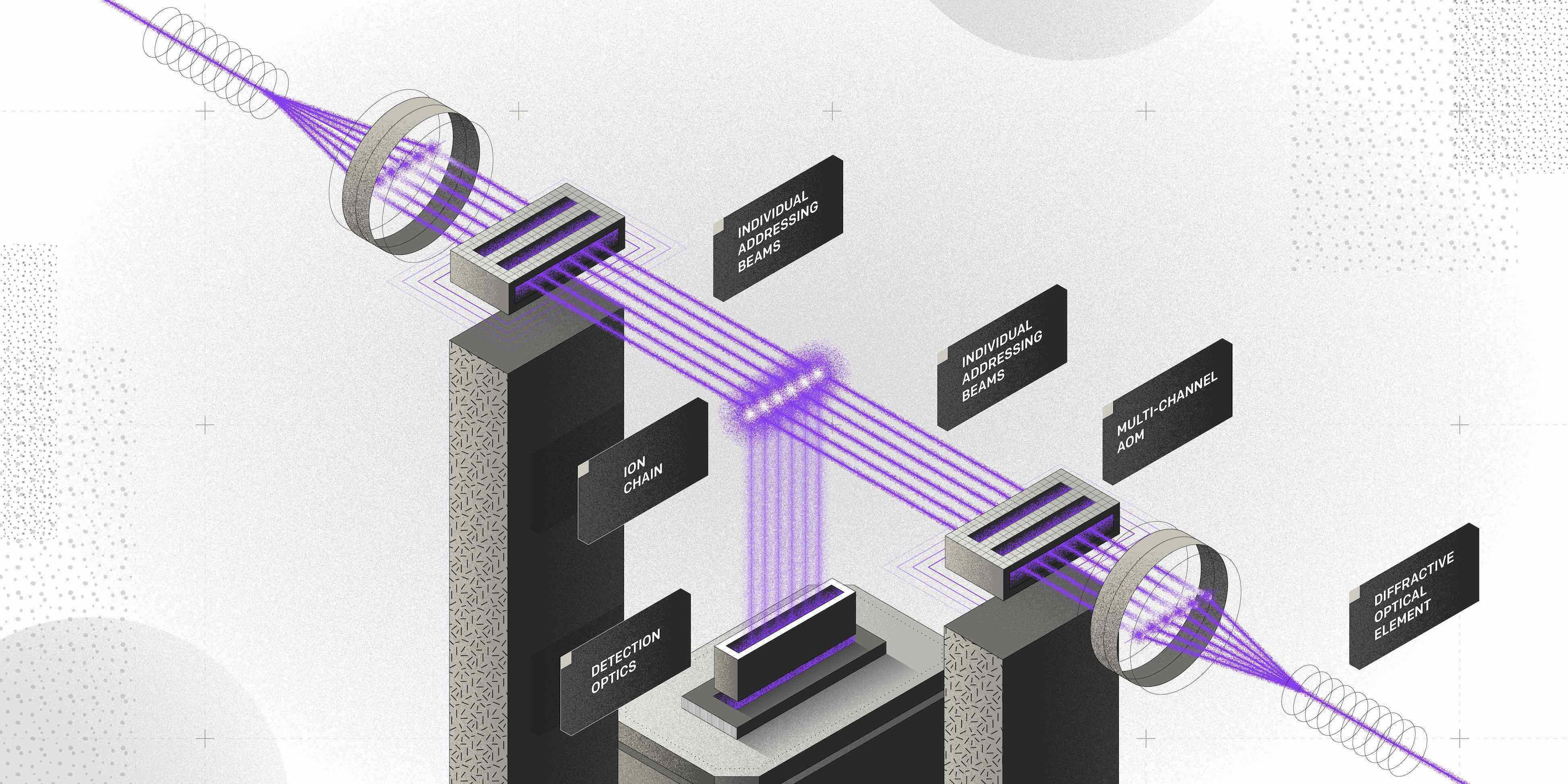}
    \caption{Schematic of next generation IonQ hardware, where two fans of individually addressable beams illuminate a chain of individually imaged ions.}
    \label{fig:oracle}
\end{figure}

\section{Maximum Likelihood Estimation based Amplitude Estimation} 

The maximum likelihood estimation (MLE) based Amplitude Estimation (AE) algorithms combine information from samples taken from quantum circuits of the form $U^t = (\mathcal{A} S_0 \mathcal{A^\dagger} S_\chi)^t \mathcal{A}$ at different depths $t$ in a statistically efficient manner, in order to provide an accurate estimate of the parameter $\theta$ of the evaluation oracle $\mathcal{A}$, where $\mathcal{A}\ket{0^k} =  \sin(\theta) \ket{1, x} + \cos (\theta) \ket{0, x'} $. An MLE based 
algorithm is specified by a schedule of measurements $(t, N_{t})$ where $N_{t}$ is the number of measurements 
made at depth $t$. A uniform (or more generally a beta distributed) prior on the success probabilities is updated according to the maximum likelihood rule applied to the results of the measurements for the different depth circuits. 
Different schedules for MLE based AE algorithms yield different tradeoffs between the number of oracle calls and the accuracy of the 
algorithm, the tradeoff can be analyzed using the notion of Fisher information of the schedule. 
Suzuki et. al \cite{S20} proposed exponential schedules $(2^{t}, N_{shot})$ where samples are taken at depths that are 
multiples of powers of $2$ ranging from $1$ up to $2^{\log (1/\epsilon)}$ and showed that such exponential schedules can be used for amplitude estimation without the need of a Quantum Fourier Transform. 
Linear and super polynomial schedules $(t^{\alpha}, N_{shot})$ for $\alpha >1$ 
have been proposed \cite{S20, TS20} along with analysis supporting the robustness of these schedules against depolarizing noise. More generally, power law schedules $(t^{\alpha}, N_{shot})$ for $\alpha >0$ were proposed in \cite{GKLPZ20} with optimal tradeoffs between circuit depth and total number of oracle calls.


Let us now describe the schedules we used in the experiments.
First, we used linear schedules, since for small depths they provide a good and simple alternative. Furthermore, for MLE based AE algorithms in a setting with depolarizing noise, it is known that the accuracy of the estimation improves only up to a certain depth, depending on the noise rate \cite{TS20, GKLPZ20}. Taking further samples at greater depths actually degrades the performance of the algorithm since it basically adds more noise to the estimate. 
Thus, taking advantage of the high fidelities of the state-of-the-art IonQ quantum computer we managed to run the amplitude estimation algorithm with depths ranging from zero to seven, where the circuit $U^7$ has ninety two two-qubit gates and a depth of sixty two, and keeping the additive error for the estimate very small, in fact below $0.02$. The schedules used $500$ shots per depth.



\begin{algorithm}[b]
\caption{Maximum Likelihood Estimation (MLE) based AE algorithm}
\begin{algorithmic}[1]
\REQUIRE Parameter $\epsilon$, maximum depth $D$ and number of shots $N_{shot}$. Access to a unitary $\mathcal{A}$ such that $\mathcal{A}\ket{0} = \sin (\theta) \ket{1}\ket{000} + \sin(\theta) \ket{0}\ket{000^{\perp}}$ and to $U^d = (\mathcal{A} S_0 \mathcal{A^\dagger} S_\chi)^d \mathcal{A}$ for  $d \in [D]$.
\STATE  Initialize the prior to the uniform distribution on angles $\theta = \frac{\pi k \epsilon}{2}$ for integer valued $k \in [0, \frac{1}{\epsilon})$, i.e. $p(\theta)= \frac{1}{\epsilon}, \forall \theta$. 
\FOR{t=1 \TO $D$ }
\STATE Initialize $N_{d_0}= N_{d_{1}}=0$, these variables record the $0$ and $1$ counts for the depth-$d$ measurements. 
\FOR{i=1 \TO $N_{shot}$}
\STATE Apply the unitary $U^d$ and measure the resulting quantum state in the standard basis. 
\STATE If the outcome is the state $\ket{1000}$ then $N_{d_0}=N_{d_0}+1$, else if the outcome is a different unary string then $N_{d_1}=N_{d_1}+1$, else do nothing.
\ENDFOR
\STATE Perform Bayesian updates $p(\theta) \to p(\theta) \cos((2d+1) \theta)^{N_{d_0}}  \sin((2d+1) \theta)^{N_{d_1}} $ and renormalize to obtain the posterior probability distribution.
\ENDFOR
\STATE Output $\theta$ with the highest probability according to the posterior probability distribution. 
\end{algorithmic}
\label{AE:mle} 
\end{algorithm}

In addition, efficient error mitigation was performed in the measurement results as in \cite{johri2021nearest} by taking advantage of the unary encoding.  

The overall algorithm used is presented in \ref{AE:mle}, it is an MLE based AE algorithm with a linear schedule $(t, 500)$ with maximum depth equal to seven and parameter $\epsilon=0.001$. The algorithm uses $1/\epsilon=1000$ buckets for the  MLE estimation, $\epsilon$ is a lower bound on the accuracy for the algorithm. The evaluation oracle used was an inner product estimation circuit for 4-dimensional unit vectors. 

The goal for the experiments was to establish an advantage for MLE based AE implementation in a noisy setting, that is to show that this family of algorithms can achieve more accurate estimates compared to the baseline of simply sampling from the evaluation oracle (the circuits corresponding to $U^0$), which is akin to classical sampling in the noiseless setting. In the noisy setting, the mean accuracy for sampling from the evaluation oracle saturates at some threshold determined by the noise rate, and thus, an AE algorithm in the noisy setting achieves an advantage if it can obtain more accurate estimates by sampling from higher depth circuits. 

\begin{figure}[t] 
\includegraphics[scale=0.5]{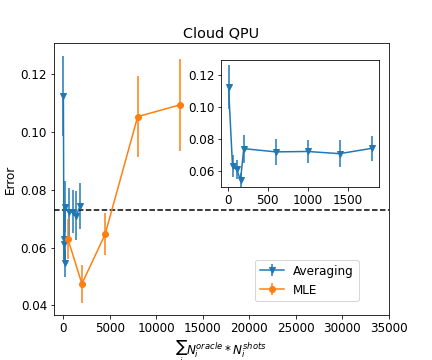}
\includegraphics[scale=0.5]{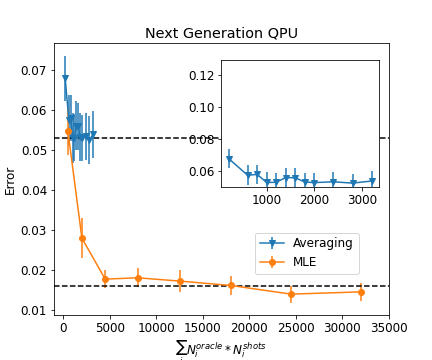}
\caption{Comparison of MLE based AE algorithm vs. simple averaging over samples from the evaluation oracle, namely the inner product estimation circuit. The mean error and error bars for MLE based AE are shown with orange circles and for simple averaging over many shots with blue triangles. The top figure corresponds to the cloud based QPU and the bottom figure to the next generation QPU with $N_{shot}=500$. The black lines are a guide to the eye. The $x$-axis denotes the total oracle calls. For simple averaging, $N^{oracle}=1$ for all points. The insets zoom in on the error from simple averaging showing its convergence to a constant value as the number of shots increases. For convenience of averaging over several circuits at each data point, the number of shots used is the total number of shots taken for each experiment even though a symmetry based postprocessing technique was utilized which discards some of the shots.}
\label{fig:MLE} 
\end{figure}

The results for the MLE based AE algorithm are presented in Figure \ref{fig:MLE}. 

The plot compares the results of estimating the angle of the evaluation oracle when we only sample from the evaluation oracle for different number of shots (blue line) versus when we perform an MLE based AE algorithm using a fixed number of shots for different circuit depths. The results are averaged over a set of 50 randomly chosen input vectors, 500 shots per circuit, and the mean and standard deviation are shown. For the top plot, the IonQ cloud-based QPU, while for the bottom plot, we used the state-of-the-art next generation IonQ QPU.

The blue lines correspond to the classical case where in a noiseless setting $O(1/\epsilon^{2})$ samples are needed to achieve accuracy $\epsilon$. Since each point is averaged over many circuits with the same structure but the same angle, depolarizing noise can be used for effective error model. Then, given the fact that there effectively exists a level of depolarizing noise even at the level of single oracle circuits, the error does not converge to zero but saturates at a higher threshold, which in the experiments (and after the post-processing technique  described in step 6 of Algorithm \ref{AE:mle}) is about $0.073$ for the cloud QPU and $0.053$ for the next generation QPU.

The orange lines correspond to the MLE based AE algorithm which uses samples from different depths and combines the estimates via an MLE calculation as in Algorithm \ref{AE:mle}). With the cloud-based QPU, we see that the MLE based algorithm beats the simple algorithm of sampling only from the evaluation oracle for depths up to two. In fact, the minimum error achieved is about $0.048$ for the MLE algorithm with maximum depth one, namely three sequential oracle applications. As we increase the depth further, samples are detrimental to the accuracy and make the approximation error shoot up. This is due to the effective depolarizing error that results from noise in the QPU operations.

On the other hand, the results on the new next generation IonQ QPU show remarkable improvements, allowing us to provide estimates with much smaller additive error for much larger depth circuits. The minimum approximation error is $0.0138$ and is achieved for depth six circuits. The improved gate fidelity for the next generation system accounts for the stark differences in the two plots and the improvement for the AE results.

Let us give some more details now about the depolarizing error model and relate it to the results of the hardware demonstration. 
Let $p= \sin^{2} (\theta)$ be the quantity being estimated by the AE algorithm. There are two different sources of effective depolarizing noise for this experiment. The combined readout and initial state preparation error $\beta$ is independent of the circuit depth. The gate error $\alpha_{t}$ is the probability of depolarizing noise due to the loss of coherence for gates in the depth $t$ evaluation. 
The combined effect of these errors is given as, 
\al{ 
\eta_{t} = \beta + \alpha_{t} 
} 
The above equation approximates $(1-\alpha_{t})(1-\beta)$ by $1-\alpha_{t} - \beta$, this is a good approximation in the regime when $\alpha_{t}, \beta$ are small.
For the depth $t$ experiment, the outcome $\ket{0}$ is observed with probability, 
$\overline{p_{t}} = p_{t} (1 - \eta_{t} ) + \eta_{t}/2 $. 

The difference between actual and estimated probability is given by, 
\al{ 
|\overline{p_{t}}  - p_{t}| = \eta_{t} | 1/2 - p_{t} | 
} 
The above equation shows that the mean error for the experiment is non zero, more precisely the mean error can be computed as the integral  $\eta_{t} \int | 1/2 - p_{t} | dp $ over 
the distribution from which the probabilities are sampled. This allows us to extrapolate the blue lines in Fig \ref{fig:MLE} and conclude that the mean error for sampling from the evaluation oracle saturates and does not improve by taking even more samples from the evaluation oracle. 
\begin{figure}[h] 
\includegraphics[scale=0.45] {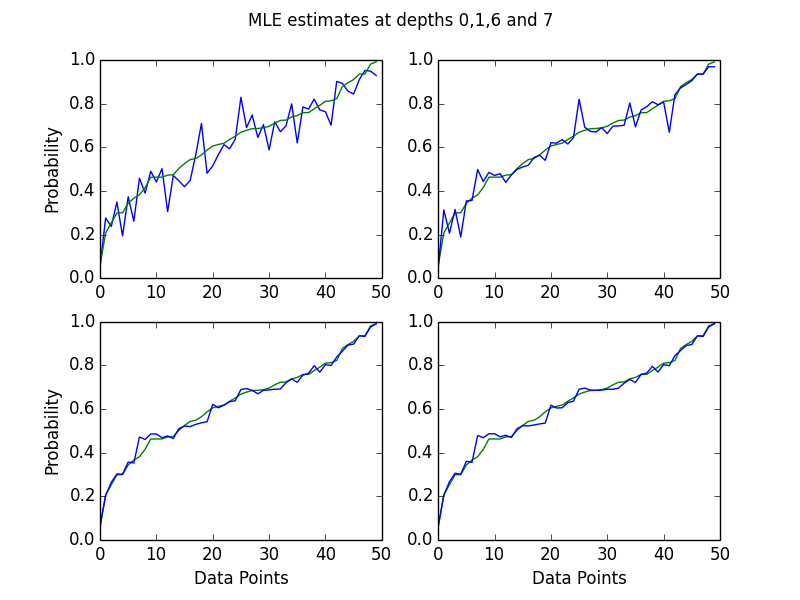} 
\caption{The green curves represent the true probabilities for the data points and the blue curves correspond to the MLE based estimates at depths $0$ (top left), 1 (top right), 6 (bottom left) and $7$ (bottom right), where the depth 6 provides the minimum mean error.} 
\label{fig:fit} 
\end{figure}

The orange line in Fig \ref{fig:MLE}  corresponds to the MLE based AE algorithm that cases achieves substantially better estimation accuracy compared to taking samples only from the evaluation oracle for the next generation QPU experiments. 
As noisier samples at higher depths are taken into account, the accuracy for the MLE based algorithm worsens for the cloud based  QPU. The lower noise rates for the next generation QPU enable the algorithm to avoid this degradation in performance for up to circuits with depth seven, that consist of more than ninety two-qubit gates and two-qubit depth of more than sixty.

In Figure \ref{fig:fit}, we plot the curves for the actual probabilities (green) and the MLE estimates (blue) at depths zero, one, six and seven, giving a global view of the results produced by the MLE based AE algorithm. The minimum mean error as metric were were obtained for MLE on circuits if depth up to six, the depth six circuits had eighty two-qubit gates and depth fifty four. The results can be viewed as a demonstration that MLE based AE algorithms can achieve improvements in accuracy compared to sampling from the evaluation oracle for near term quantum devices in a noisy setting.

\section{Chinese Remainder Theorem based Amplitude Estimation algorithms} 

In this section, we look at a different type of low depth AE algorithm that uses the Chinese Remainder Theorem instead of MLE techniques to combine the samples at different circuit depths. This algorithm, called QoPrime has a rigorous proof of convergence and in the noiseless case or in the case of depolarizing noise applied independently over the samples, needs fewer samples than the MLE based algorithm with linear schedules and matches the one with power law schedules \cite{GKLPZ20}. 

For the experiment, we use a simplified variant of the QoPrime algorithm that samples at two consecutive depths to find low precision estimates for the amplitude that are then combined using the Chinese remainder theorem to provide better estimates. 
Given the fact that we choose two consecutive odd numbers as moduli, the rounding and sign estimation procedures are also simplified compared to the general algorithm in \cite{GKLPZ20}, where here we will use an MLE based estimate for very small depths (up to depth two) as a way to find the correct sign.

\begin{algorithm}
\caption{Chinese Remainder Theorem (CRT) based AE algorithm.}
\begin{algorithmic}[1]
\REQUIRE 
Accuracy parameter $\epsilon$, maximum depth $D$ and number of shots $N_{shot}$. Access to a unitary $\mathcal{A}$ such that $\mathcal{A}\ket{0} = \sin (\theta) \ket{1}\ket{000} + \sin(\theta) \ket{0}\ket{000^{\perp}}$ and to $U^t = (\mathcal{A} S_0 \mathcal{A^\dagger} S_\chi)^t \mathcal{A}$ for  $t \in [D]$.
\STATE Compute an MLE estimate $p_{0}$ for the quantity $\sin^2(\theta)$ using Algorithm IV.1 with maximum depth equal to $2$ and define angle $\theta' = \arcsin(\sqrt{p_{0}})$. 
\STATE Compute estimates $p_{D}$ and $p_{D-1}$ of the quantities $\sin^2 ((2D+1)\theta)$ and $\sin^2 ((2D-1)\theta)$ by using samples from the circuits $U^D$ and $U^{D-1}$ respectively. 
\STATE Compute $l = \frac{2(2D-1)}{\pi} \arcsin(\sqrt{p_{D}})$ and $h =  \frac{2(2D+1)}{\pi} \arcsin(\sqrt{p_{D-1}})$. 

\STATE Compute $s_{1} = sgn( \sin (2(2D-1))\theta' )$ and  $s_{2} = sgn( \sin (2(2D+1))\theta' )$. 

\STATE Compute integers $v_{t} \mod (4D^{2}-1)$ for $1\leq t \leq 4$ such that $v_{t} \mod (2D-1)= \lfloor s1.l\rfloor  + (t\mod 2)$ and $v_{t} \mod (2D+1)= \lfloor s2.h \rfloor + t/2$ using the 
Chinese remainder theorem. 
\STATE Out of the 4 values computed in step 4, select $v_{t}$ minimizing $|\sin^{2}(\frac{v_{t}\pi}{4D^{2} -1}) - p_{0}|$. 
\STATE 
Output the value $\frac{v_{t} \pi}{4D^{2} -1}$ as estimate for the angle $\theta$. 
\end{algorithmic}
\label{AE:crt}
\end{algorithm}
The main idea for the CRT based AE Algorithm \ref{AE:crt} is that if estimates obtained from running the evaluation oracle at depths $D$ and $D-1$ are sufficiently accurate, then they can be combined to get a higher accuracy estimate using the Chinese remainder theorem. 
The depth $D$ and $D-1$ estimates are determined up to an ambiguity in sign which are resolved using an approximation to the true success probability obtained with a depth 2 MLE estimate. The CRT based AE Algorithm \ref{AE:crt} requires fewer oracle calls than the MLE algorithm as it considers samples only at depths $0,1,2$ and at $D, D-1$.

\begin{figure} [b] 
\includegraphics[scale=0.32]{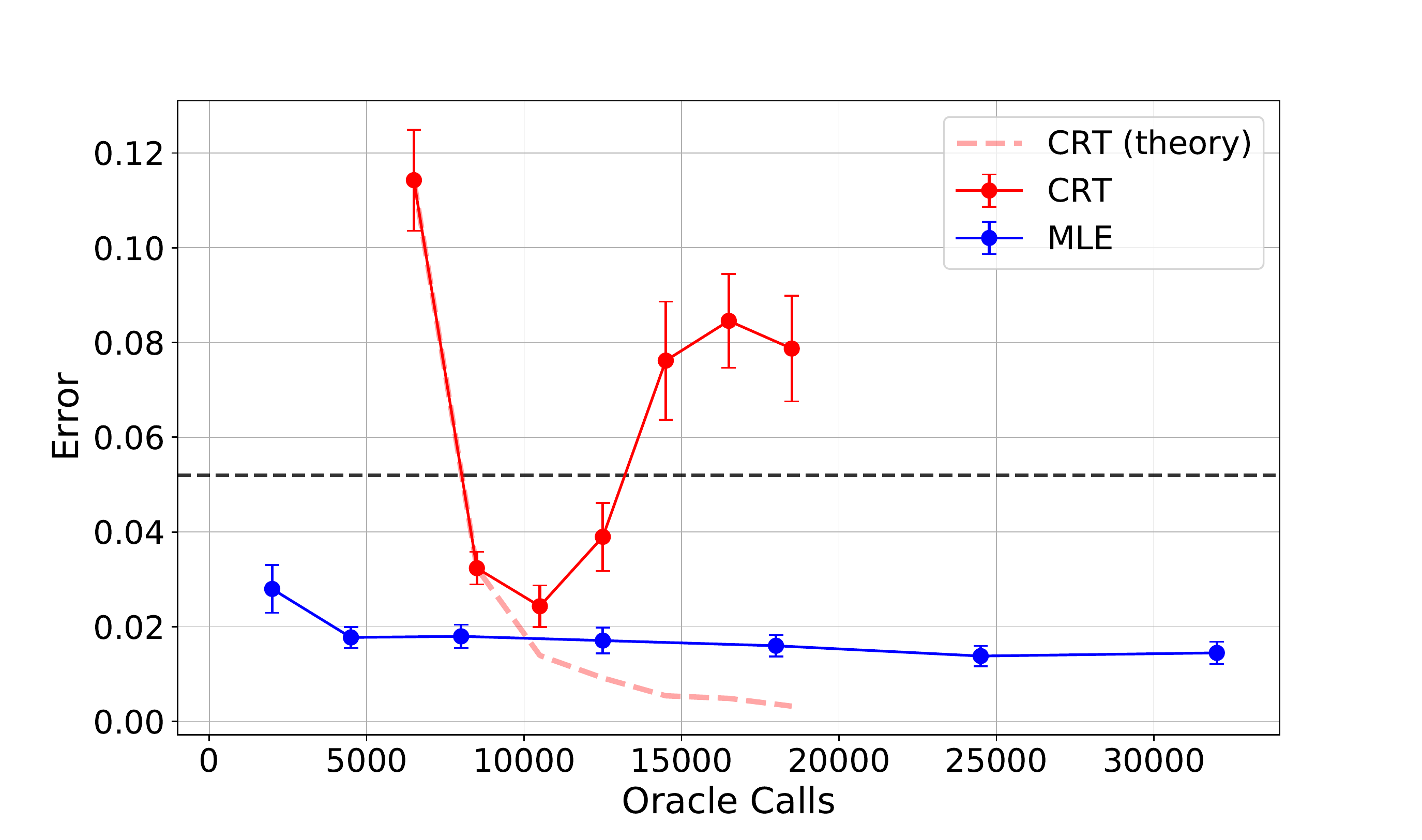} 
\caption{Comparison of mean error against total number of oracle calls for MLE, exact CRT and CRT algorithms for depths ranging from 1 to 7. The black dashed line represents the baseline error for sampling from the evaluation oracle,  same as the black line in Fig \ref{fig:MLE}. } 
\label{fig:CRT} 
\end{figure} 

Figure \ref{fig:CRT} shows the results of Algorithm \ref{AE:crt} with the inner product estimation circuit as the evaluation oracle on the IonQ next generation QPU for depths ranging from one to seven. As we said, here the CRT based algorithm of depth $D$ takes samples only from circuits of depths $D$ and $D-1$, while it also uses an MLE-based algorithm for depth two to differentiate the signs of the estimates. Note as well that the cloud QPU was not able to show that this algorithm improves the approximation error with respect to sampling directly from the evaluation oracle, and the next generation QPU was necessary to show such an improvement.

In Figure \ref{fig:CRT}, the green curve represents the error if the exact success probabilities $\sin^{2}((2D-1) \theta)$ and $\sin^{2}((2D+1) \theta)$ are used instead of the estimated probabilities in step 2 of the CRT algorithm. Using the exact success probabilities instead of the estimates obtained from depth $D$ and $D-1$ runs of the oracle corresponds to the accuracy of the Algorithm \ref{AE:crt} for a noiseless setting where the number of shots is large enough to ensure convergence to the true success probabilities. The red curve represents the accuracy of the CRT algorithm \ref{AE:crt}. This achieves a minimum mean error of 0.024 at depth three. The figure shows that the errors for the CRT algorithm \ref{AE:crt} match the errors for the noiseless algorithm for depths up to two corresponding to the co-prime moduli $(3,5)$. The CRT algorithm is less robust against noise compared to the MLE as it requires both probabilities $p_{D}$ and $p_{D-1}$ to be good approximations to the true probabilities. The result for the CRT will be incorrect if either of these probabilities is not estimated correctly.

We also remark that while the MLE based algorithm performs better on average than the CRT based algorithm, the  noiseless success probabilities for the CRT algorithm represented by the green line are lower than the empirically observed probabilities for the MLE based algorithm in Figure \ref{fig:MLE}.

\begin{figure} [h]
\includegraphics[scale=0.45]  {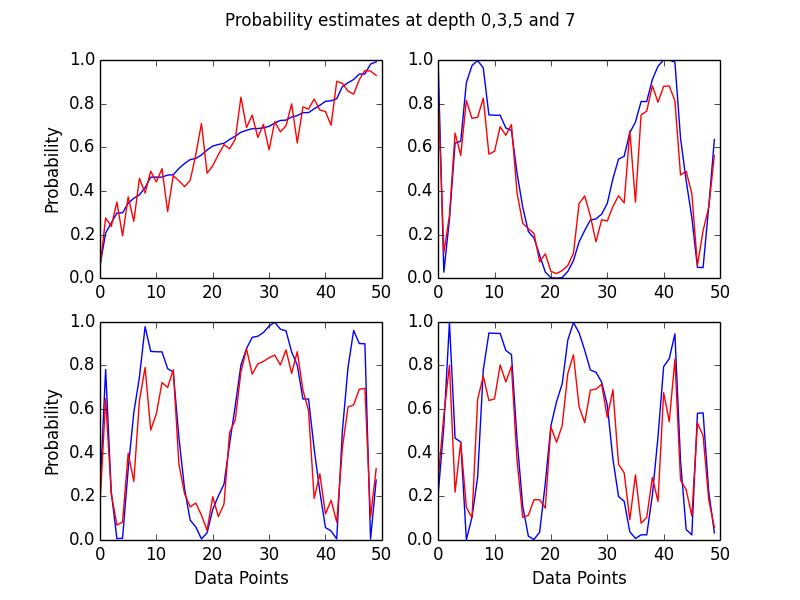} 
\caption{Exact probabilities $\sin^{2}((2D+1)\theta)$ and the probability estimates with 500 shots for the depths $0$ (top left), 3 (top right), 5 (bottom left) and $7$ (bottom right). } 
\label{crt:fit} 
\end{figure} 

Let us try to understand in more detail the performance of the CRT based algorithm. Incorrect estimates at higher depths are more common due to greater noise for deeper circuits, resulting in high average error for the CRT algorithm for samples of depth greater than three. Figure \ref{crt:fit} plots the true success probabilities at depths zero, three, five, and seven (blue) and the estimates from the runs of the corresponding oracles with 500 shots (red). It can be seen from the figures that the AE estimates are not always a close approximation for the true success probabilities. The inputs to the CRT procedure computed in step 3 of the algorithm differ from the correct estimates for the outlying data points resulting in an error in the final answer. This results in the CRT algorithm having higher mean error than the MLE algorithm. 

We can also relate the error models and the experimental results of the CRT based algorithm.
The theoretical analysis of the QoPrime algorithm assumed independence of errors over samples so that Chernoff bounds could be applied. However, in the hardware implementation the errors over the 500 shots are not independent but correlated in time with larger errors sometimes occurring on consecutive runs of the circuit, and this results in a non-smooth approximation to the success probabilities. It should be noted that the next generation QPU is an experimental system and further improvements in gate fidelity and calibration should improve the quality of approximation for the data points and the performance of the CRT based AE algorithm.

\section{Amplitude Estimation with experimental hyperparameter tuning}

In this section we show that we can further improve the performance of the amplitude estimation algorithms by using experimental results on the quantum hardware to tune some hyper parameters of the algorithms.

\subsection{Hybrid Amplitude Estimation algorithm}

We start with Algorithm \ref{AE:hybrid}, a simple hybrid algorithm that compares between a low depth MLE based estimate and the CRT based estimate and tries to output the one with the smaller error. 
Such an algorithm tries to use the higher CRT estimates when they are correct, but falls back to an MLE based estimate when the CRT estimate seems to be wrong. In order to choose between the CRT and MLE based estimates we cannot use the estimate with the real smaller error as the true success probabilities are unknown. Instead, we use a heuristic test condition that outputs the CRT estimate if the MLE and CRT estimates do not differ more than a $\beta$ factor times some average difference $|MLE_{Avg}(2) - CRT_{exact}(D)|$ between the two estimates which are quantities that can be computed through experiments on some training data.  

More precisely, in order to define these quantities we fix a prior distribution on probabilities, this can be the uniform or a beta distributed prior for example. $MLE_{Avg}(2)$ is the average mean error for the MLE algorithm at depth 2 (or some other low depth) with the probabilities drawn from the prior distribution. $CRT_{exact}(D)$ are the mean errors for the classically simulated CRT algorithm with probabilities drawn from the prior distribution. In this case, they can be read off from the blue and green lines as in Fig \ref{fig:CRT}. The factor $\beta$ can also be tuned experimentally.

\begin{algorithm}[H]
\caption{Hybrid AE algorithm.}
\begin{algorithmic}[1]
\REQUIRE  Accuracy parameter $\epsilon$, parameter $\beta$, maximum depth $D$ and number of shots $N_{shot}$. Access to a unitary $\mathcal{A}$ such that $\mathcal{A}\ket{0} = \sin (\theta) \ket{1}\ket{000} + \sin(\theta) \ket{0}\ket{000^{\perp}}$ and to $U^t = (\mathcal{A} S_0 \mathcal{A^\dagger} S_\chi)^t \mathcal{A}$ for  $t \in [D]$. 
\STATE Compute an MLE estimate $p_{0}$ for the quantity $\sin^2(\theta)$ using Algorithm IV.1 with maximum depth equal to $2$.  
\STATE Compute CRT estimate $q_{0}$ with maximum  depth $D$ using the CRT algorithm \ref{AE:crt}. 
\STATE If $|p_{0} - q_{0}| > \beta |MLE_{Avg}(2) - CRT_{exact}(D)|$ return $p_{0}$
else return $q_{0}$.
\end{algorithmic}
\label{AE:hybrid} 
\end{algorithm}

\begin{figure} [t]
    \centering
    \subfloat[\centering Depth 5]{{\includegraphics[scale=0.4]{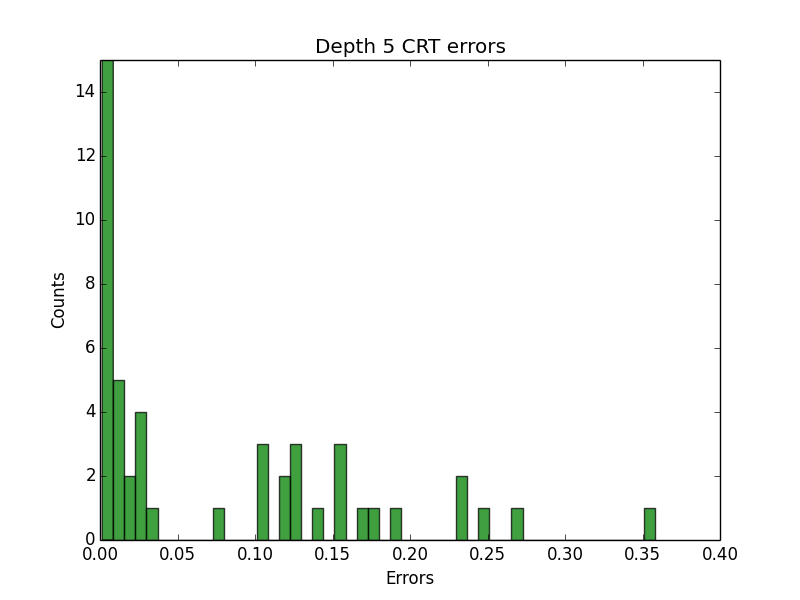} }}%
    \qquad
    \subfloat[\centering Depth 7]{{\includegraphics[scale=0.4]{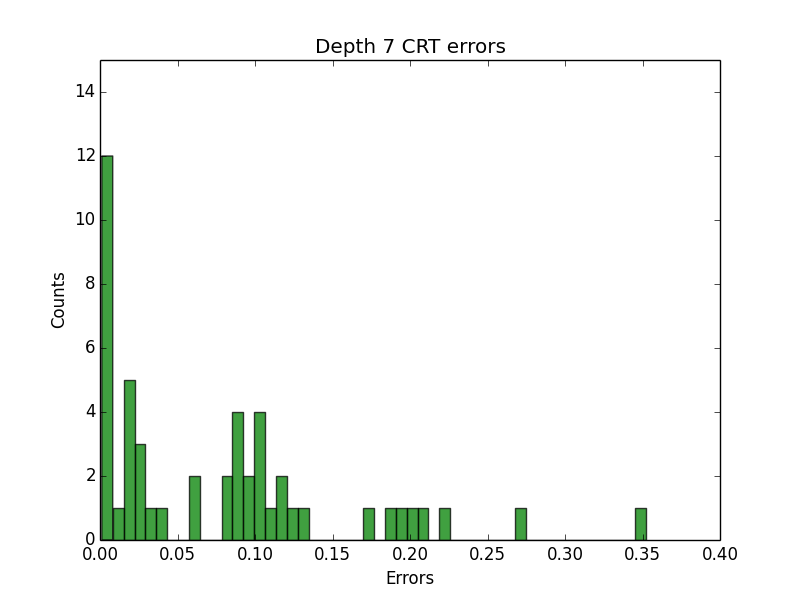} }}%
    \caption{Histograms for errors for the CRT based AE algorithm at depths 5 and 7. This shows a distinct {\em failure mode} for the CRT algorithm, in that many of the points are estimated accurately, while others are entirely miscalculated by the algebraic reconstruction of the CRT. The hybrid algorithm \ref{AE:hybrid} is designed to diagnose this issue.}%
    \label{fig:hist}%
\end{figure}

Another way to see the potential advantage of the CRT based approach is to look at the histograms for the errors achieved by the algorithm at high depths. Figure \ref{fig:hist} shows the error histograms for the CRT based algorithm with maximum depth five and seven. It can be seen from the figure that even at high depths, the CRT based AE algorithm is able estimate a large fraction of the points with very low errors. The larger mean errors compared to MLE can be explained by a few outlying points with large errors. The algorithm is close to the theoretical accuracy for depths up to three.

The performance of the hybrid AE algorithm compared to MLE and the CRT is illustrated in Figure \ref{fig:all}. The estimates output by the hybrid algorithm are considerably better than the CRT based ones and match the MLE estimates for up to depth equal to five. The hybrid algorithm achieves a minimum mean error of 0.017 at depth four, which improves upon the mean error $0.018$ for the depth two MLE algorithm. Note also that the hybrid algorithm uses fewer oracles calls than the MLE based algorithm for the same maximum depth.


\subsection{Power law Amplitude Estimation}

    

The MLE based amplitude estimation algorithm \ref{AE:mle} used a linear schedule and a fixed number of shots at each depth. Namely, we used a schedule of the form $(d, N_{shot})$ for $d$ all integers up to $D$, the maximum depth. Here we introduce an additional degree of freedom in the choice of the exponents in the {\em power-law AE algorithm}, studied in \cite{GKLPZ20}. We propose a noise aware power law AE method that solves an optimization problem to determine the optimal exponent for a given target error. The sampling schedules will be of the form:
 \begin{equation}
    \label{eq:powerlaw}
     N_d = \left\lfloor N_\text{shots} \times (2d+1)^\nu\right\rfloor,
\end{equation}
where $N_d$ is the number of samples taken at depth $d$, for $d=0,1,2,3,\ldots$. The number of shots $N_\text{shots}$ was taken to be $500$ and the exponent $\nu$ is the hyper-parameter of the algorithm that we tune experimentally. 
Assuming an unbiased estimator, the Cram\'{e}r-Rao bound can be used to connect the schedule to the estimation error. 
When there is a specific level of (depolarizing) noise $\gamma_d$ associated with depth $d$, the classical Fisher information is exponentially damped due to the noise. Given the sampling schedule in \eqref{eq:powerlaw}, the problem becomes finding the exponent $\nu$ that minimizes the oracle calls for given a desired target error, namely:
\begin{align}
    \label{eq:powerlawoptimization}
    \text{minimize}_{\nu} & \;\;\;  \mathcal O \sim \sum_{d=0}^D (2d+1)^{\nu + 1} 
\end{align}
\begin{align}
    \label{fisher}
    \text{subject to} & \;\;\; \mathcal 
    F_\text{noisy}\equiv N_\text{shots}\sum_{d=0}^D(2d+1)^{\nu + 2}e^{-2\gamma_d}\geq \epsilon^{-2}
\end{align}
where we introduce a parameter-agnostic approximation $\mathcal F_\text{noisy}$ to the true Fisher information. This optimization problem is solved classically, given experimental estimates of the noise levels $\gamma_d$, and allowing the exponent $\nu$ to take values smaller than $-1$. For the data under consideration, a depolarizing model was fit with noise parameter $\gamma_d$ ranging from $\gamma_0=0.035$ at depth zero (evaluation oracle), up to $\gamma_7=0.35$ at depth seven. 

The last step of the algorithm is similar to the first MLE method in \ref{AE:mle} in that it performs a noise-aware Bayesian update on the parameter $\theta$ given the data at each depth $d$:
\begin{equation}
        \label{eq:noiseawarebayesianupdate}
         p(\theta\,|\,N_{d_0},\,N_{d_1}) \sim p_d(0)^{N_{d_0}} p_d(1)^{N_{d_1}}\,p(\theta)
\end{equation}
Where $N_{d_0}$ and $N_{d_1}$ are the number of $0$s and $1$s measured at depth $d$, and in the depolarizing noise assumption the measurement probabilities are:
\begin{equation}
    \label{eq:depolarizingmodel}
    p_d(1) = \frac{1 - e^{-\gamma_d}\cos(2(2d+1)\theta)}{2},\quad p_d(0) = 1 - p_d(1)
\end{equation}

The maximum-likelihood estimates for the power-law method were performed on data which is sampled without replacement from the measurements at each depth. Namely, for every target error we construct a dataset with $N_d\leq N_0=500$ Bernoulli trials for depths $d\leq 7$, drawn from the dataset of all measurements.

Finally, one could question the use of the power-law ansatz \eqref{eq:powerlaw} in the first place, when the optimization problem in \eqref{eq:powerlawoptimization} could be formulated more fundamentally as a linear program over the sampling schedule $N_d$. The answer lies in the statistical subtleties which appear when translating the amplitude estimation problem into the language of classical inference. Specifically, the simple schedules arising out of solving a na\"ive linear program would most likely {\em not lead to unbiased estimators} due to the periodic nature of the likelihood function, as pointed out in \cite{GKLPZ20}. In short, the choice of a power law schedule is an empirically-validated compromise which ensures an unbiased estimator, and thus legitimizes the use of the simple Cram\'{e}r-Rao bound above. More work is needed in order to obtain provable guarantees with such sampling schedules.

\begin{algorithm}[H]
\caption{Noise aware Power Law MLE algorithm.}
\begin{algorithmic}[1]
\REQUIRE  
 Target accuracy $\epsilon$, maximum depth $D$, number of shots $N_\text{shots}$, noise levels $\gamma_d$ for $d \in [D]$.
Access to a unitary $\mathcal{A}$ such that $\mathcal{A}\ket{0} = \sin (\theta) \ket{1}\ket{000} + \sin(\theta) \ket{0}\ket{000^{\perp}}$ and to $U^d = (\mathcal{A} S_0 \mathcal{A^\dagger} S_\chi)^d \mathcal{A}$ for  $d \in [D]$. 
\STATE Find optimal power-law schedule $S=\{ (d,N_d) \}$, with $d \in [D]$, using Equations \eqref{eq:powerlaw} and \eqref{eq:powerlawoptimization}.
\STATE  Initialize the prior to the uniform distribution on angles $\theta = \frac{\pi k \epsilon}{2}$ for integer valued $k \in [0, \frac{1}{\epsilon})$, i.e. $p(\theta)= \frac{1}{\epsilon}, \forall \theta$. 
\FOR{$(d,N_d) \in S$}
\STATE Initialize $N_{d_0}= N_{d_{1}}=0$, these variables record the $0$ and $1$ counts for the depth-$d$ measurements. 
\FOR{i=1 \TO $N_{d}$}
\STATE Apply the unitary $U^d$ and measure the resulting quantum state in the standard basis. 
\STATE {\em Postprocessing:} If the outcome is the state $\ket{1000}$ then $N_{d_0}=N_{d_0}+1$, else if the outcome is a different unary string then $N_{d_1}=N_{d_1}+1$, else do nothing.
\ENDFOR
\STATE Perform Bayesian updates according to \eqref{eq:noiseawarebayesianupdate} and renormalize to obtain the posterior probability distribution.
\ENDFOR
\STATE Output $\theta$ with the highest probability according to the final posterior probability distribution. 

\end{algorithmic}
\label{AE:hybrid} 
\end{algorithm}

The performance of the power law AE algorithm compared to previous AE algorithms is illustrated in Figure \ref{fig:all}. The power law algorithm was performed for ten different target errors $\epsilon$, where for each case a new optimization problem was solved in order to find the optimal schedules for the algorithm. We also plot the theoretical curve that corresponds to Eq. \ref{fisher}.    


\section{Comparison of Amplitude Estimation algorithms}

\begin{figure}[t]
    \centering
    \includegraphics[scale=0.29]{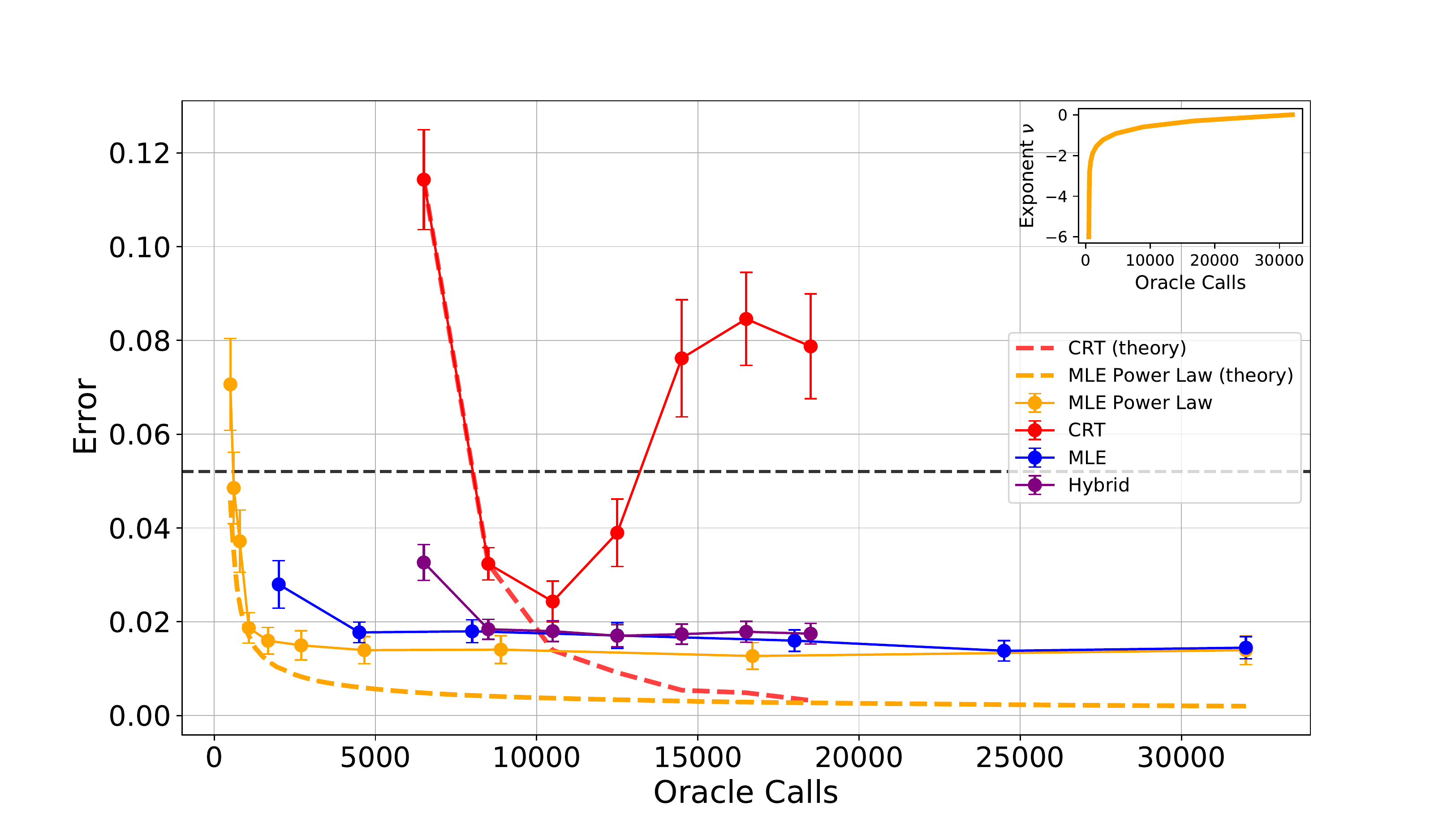}
    \caption{Comparison of the all the amplitude estimation algorithms discussed in this work. The maximum-likelihood methods perform the best, particularly when the schedules are optimized with the power-law method (orange). The CRT based algorithms have higher mean error, despite the theoretical guarantees of the algorithm. The black dashed line represents the baseline mean error for sampling from the depth $0$ evaluation oracle. The inset shows the optimal exponent $\nu$ chosen at various error levels by the power-law algorithm.
    }
    \label{fig:all}
\end{figure}

In Figure \ref{fig:all} we provide a comparison between all different amplitude estimation methods we described above, namely sampling from the evaluation oracle, MLE based with linear schedules, MLE based with power law schedules, CRT based, and hybrid. We see that the MLE based algorithms perform better on hardware, and the one using power law schedules with  exponents chosen experimentally achieves the best performance. The CRT based algorithms have higher mean error than the MLE based algorithm, though they have better theoretical guarantees and provide good estimates for many inputs. The black dashed line represents the baseline error for sampling from the evaluation oracle.

\section{Conclusions} 
We reported the results for experiments on low depth Amplitude Estimation algorithms using MLE and CRT based approaches on a state-of-the-art trapped ion quantum computer. The high fidelity of the quantum hardware allowed us to run oracle circuits with depths ranging up to seven which translates to four qubit circuits with more than ninety two-qubit gates and depth sixty. Similar experiments on quantum hardware available on the cloud provided considerably worse results.

The MLE based algorithms showed significant improvements in accuracy when higher depth samples were taken into account reaching errors of less than $0.014$ at depth six, while the error when samples from the evaluation oracle are taken saturates to $0.053$. We also developed a more sophisticated version of maximum-likelihood amplitude estimation based on a power-law schedule. This introduced two improvements: first, the asymptotic precision improved since the power-law algorithm incorporates a noise model, albeit an imperfect one. Second, this noise floor is reached much faster in terms of oracle calls since the optimal power-law schedules spend fewer shots at costly higher depths. Note that all maximum-likelihood methods can naturally accommodate any probabilistic noise model in the definition of the likelihoods.

The CRT based algorithm is more sensitive to noise and it was affected by the hardware noise as well as the correlated errors across experiments. It achieved a minimum mean error of $0.024$ at depth 3, following its design precision curve, before departing from it at larger depths. A hybrid algorithm that combines small depth MLE estimates with CRT estimates achieved minimum mean error of $0.017$, an improvement over the depth two MLE estimates with an average error of $0.018$. With improvements in hardware fidelity and calibration the CRT based and the hybrid algorithms will become competitive with the MLE based approach. 

Note that we restricted the experiments to four qubits, because our main goal was to probe the regime where the evaluation oracle is invoked a large number of times in a noisy setting, achieving up to fifteen sequential oracle invocations with still excellent results. A next step would be to establish tradeoffs between circuit depth and number of oracle calls in an experimental setting, as theoretically proved in \cite{GKLPZ20}, and this may soon become feasible with further improvements in hardware. 

\section{Author Contribution}

A.P., T.G-T, W.Z. and I.K. designed the algorithms and performed the analysis for the experimental data. J.N., N.P., K.S., K.W. and S.J. performed the experiments on the IonQ hardware.

\bibliographystyle{plain} 
\bibliography{b1.bib}

\end{document}